\documentclass[amstex,useAMS,epsf,amssymb,usenatbib, twocolumn]{mn2e}
\usepackage{epsfig}
\usepackage{graphicx}
\usepackage{color}
\usepackage{amssymb,amsmath}
\usepackage{multicol}
\usepackage{soul}
\usepackage[normalem]{ulem}
\usepackage{url}
\usepackage{hyperref}
\usepackage{breakurl}

\usepackage{tabularx}

\newcolumntype{Y}{>{\centering\arraybackslash}X}

\newcommand{\fermi}{{\it Fermi}}
\newcommand{\fermilat}{{\it Fermi}-LAT}
\newcommand{\gray}{$\gamma$-ray}
\newcommand{\grays}{$\gamma$ rays}

\title[A temporal analysis for pair halo searches] {A search for pair halos around active
galactic nuclei through a temporal analysis of \fermilat{} data}

\author[D. A. Prokhorov \& A. Moraghan]{D. A. Prokhorov$^{1}$
\thanks{E-mail:phdmitry@gmail.com} and A. Moraghan$^{2, 3}$
\thanks{E-mail:ajm@asiaa.sinica.edu.tw}
\\
~\\
$^{1}$ Department of Physics and Electrical Engineering, Linnaeus
University, 351 95 V\"{a}xj\"{o}, Sweden
\\
$^{2}$ Academia Sinica Institute of Astronomy and Astrophysics, P.O.
Box 23-141, Taipei 106, Taiwan
\\
$^{3}$ Center for Galaxy Evolution Research and Department of
Astronomy, Yonsei University, Seoul 120-749, Republic of Korea}

\pagerange{\pageref{firstpage}--\pageref{lastpage}} \pubyear{2015}

\begin{document}

\maketitle

\begin{abstract}
We develop a method to search for pair halos around active galactic
nuclei (AGN) through a temporal analysis of \gray{} data. The basis
of our method is an analysis of the spatial distributions of photons
coming from AGN flares and from AGN quiescent states and a further
comparison of these two spatial distributions. This method can also
be used for a reconstruction of a point spread function (PSF). We
found no evidence for a pair halo component through this method by
applying it to the \fermilat{} data in the energy bands of 4.5-6,
6-10, and $>$10 GeV and set upper limits on the fraction of photons
attributable to a pair halo component. An illustration of how to
reconstruct the PSF of \fermilat{} is given. We demonstrate that
the PSF reconstructed by using this method is in good agreement with
that which was obtained by using the \gray{} data taken by LAT in
the direction of the Crab pulsar and nebula.
\end{abstract}


\section{Introduction}

The existence of possible extended, diffuse, \gray{} sources
(so-called pair halos) around active galactic nuclei (AGNs) was
predicted by \citet{Aharonian1994}. \grays{} with energies above
$\sim$1 TeV emitted by distant AGNs cannot propagate over
cosmological distances because of electron-positron pair production
($\gamma + \gamma \rightarrow \mathrm{e}^{-} + \mathrm{e}^{+}$) on
the optical/infrared extragalactic background light \citep[EBL, see,
e.g.,][]{Kneiske2002, Franceschini2008, Finke2010}. The
electron-positron pairs created in the $\gamma$-$\gamma$
interactions travel in the intragalactic magnetic field and emit
secondary cascade \grays{} with energies $\sim$1 GeV owing to
inverse Compton (IC) scattering by cosmic microwave background (CMB)
photons. Thus, a \gray{} image of an AGN is expected to exhibit a
halo of secondary photons around a central point-like source. A
potential detection of pair halos around AGNs will provide us with a
measurement of the strength of the intergalactic magnetic field
\citep[for a review, see][]{Neronov2009}. The existence of this weak
``seed'' magnetic field itself is unavoidable because this is
required for the production of magnetic fields in galaxies and in
clusters of galaxies which are about $10^{-6}$ gauss via field
amplification mechanisms \citep[for a review, see][]{Ryu2012}.

The \fermi{} Large Area Telescope (LAT) is a pair conversion
telescope designed to cover the broad energy range from 20 MeV to
greater than 300 GeV \cite[][]{Atwood2009}. The broad energy range
of \fermilat{}, therefore, covers typical energies of secondary IC
photons if the energies of primary \grays{} are between $\simeq$1
and $\simeq$20 TeV. The \fermilat{} normally operates in sky-survey
mode where the whole sky is observed every 3 hr and \fermilat{}
sensitivity allows us to monitor AGNs on a daily basis. This quality
of \fermilat{} data is particularly important to search for \gray{}
flares of AGNs. The second catalog of high-energy \gray{} sources
\citep[the 2FGL catalog,][]{Nolan2012} detected by \fermilat{} and
derived from data taken during the first 2 years of the \fermi{}
mission was released by the \fermilat{} collaboration. The 2FGL
catalog contains 1873 sources detected and characterised in the 100
MeV to 100 GeV range of which 127 are firmly identified and 1171 are
reliably associated with counterparts of known or likely \gray{}
producing source classes. Note that the 2FGL catalog contains 1047
\gray{} sources associated with AGNs, mostly blazars, and,
therefore, this source class is dominant in the catalog. Since AGNs
are extragalactic sources, AGNs are almost uniformly
distributed on the \gray{} sky. This property will allow us to
select a sample of AGNs that are not projected on the Galactic
plane and, therefore, this sample will be less contaminated by
Galactic \gray{} emission.

Pair halos have not yet been detected through a stacking analysis of
AGN images in \fermilat{} data \citep[][]{Neronov2011,
Ackermann2013}. However, an important and continuous work by
the \fermilat{} team on \gray{} event reconstruction leads to
better understanding the performance of the LAT instrument. The
continuous accumulation of the data by the LAT improves the
sensitivity in the search for pair-halos. Meanwhile, a
non-detection of pair halos has put important limits on the
intergalactic magnetic field, $B\geq10^{-16}-10^{-15}$ gauss through
non-observations of GeV emission  from TeV blazars
\citep[][]{Neronov2010, Tavecchio2010, Dolag2011}. Note that the
value of intergalactic magnetic field strength has not been measured
by any approach and that \gray{} astronomy provides us with
promising methods to measure this magnetic field strength even
possibly in the near future \citep[see][]{Neronov2009}. Apart
from a direct search of the extension of \gray{} emission
around AGNs, there are independent and complementary methods to
probe the intergalactic magnetic field using \gray{} data,
including a search for time delays in \gray{} flares
\citep[see][]{Plaga1995, Murase2008}, studies based on spectral data
alone \citep[][]{Essey2011}, or observations of astronomical
events such as solar occultations of blazar 3C 279
\citep[][]{Fairbairn2010}\footnote{Note that the Sun is a bright
extended gamma-ray source. The solar gamma-ray flux is not small
compared with that of 3C 279 (at least during the solar cycle
minimum). The emission from the solar disk is due to the cosmic-ray
cascades in the solar atmosphere and the largely extended emission
around the Sun is due to the inverse Compton scattering of
cosmic-ray electrons off solar photons \citep{solar2011}. This
along with the short duration of occultations significantly
complicate a search of a pair halo around 3C 279 during solar
occultations \citep[][]{Barbiellini2014, Kotelnikov2015}.}.

The importance of a careful point spread function (PSF)
reconstruction from the on-orbit \fermilat{} data was emphasized by
\citet[][]{Ando2010} and \citet[][]{Neronov2011}. \citet[][]{Neronov2011} compared
the stacked AGN signal to the signal of the Crab pulsar and nebula,
which is a bright galactic gamma-ray source, and found that the
shapes of the two signals coincide. This means that the entire
stacked AGN signal is well described by a point-source signal. A
detailed analysis by \citet[][]{Ackermann2013} showed that the PSF
determined from two years of on-orbit data above 3 GeV is found to
be broader than the pre-launch PSF determined through extensive
Monte Carlo simulations and beam tests (this agrees well with the
results by \citet[][]{Neronov2011}). To calibrate the PSF, they
adopted a technique of stacking sources, where the angular offsets
of \grays{} from their presumed sources are analyzed as if they came
from a single source. Pulsars would be ideal for calibrating the
PSF. However, the \grays{} from pulsars above 10 GeV are limited.
They selected a subset of 65 AGNs to calibrate the PSF. Out of the
65 sources, 35 were associated with BL Lac-type blazars, 27 with
Flat Spectrum Radio Quasars (FSRQ), 1 with a non-blazar active
galaxy, and 2 with an active galaxy of uncertain type.
\citet[][]{Ackermann2013} estimated the 68\% containment radii for
front and back events from the Geminga and Vela pulsars below 31.6
GeV and the AGN calibration data set above 3.16 GeV. Using the
pulsed \gray{} emission between 1 GeV to 31.6 GeV from pulsars,
which appears as a true point source to the \fermilat{}, they placed
limits on the angular extension of AGN emission relative to pulsar
emission and derived the upper limits for the fraction of \grays{}
from the stacked AGN sample attributable to a pair halo component.

In this paper, we develop a method to reconstruct the PSF and to
search for pair halos through a temporal analysis of AGN emission.
We illustrate its application to \fermilat{} data. The \gray{}
emission of AGN can be divided into three components, which are
\gray{} flares, quiescent \gray{} emission from a compact region
around AGNs, and extended pair halo emission. The first component is
strongly variable in time, while the pair halo component is
time-invariant. Numerous flares from AGNs were detected with
\fermilat{}
\footnote{\burl{http://fermi.gsfc.nasa.gov/ssc/data/access/lat/msl_lc/}}
\citep[][]{Abdo2010}. Thus, for example, an active galaxy 3C 454.3
located 7.2 billion light-years away, was the brightest source in
the gamma-ray sky during its strong gamma-ray flaring activity
\footnote{Most of the time, the brightest source in
the gamma-ray sky is the Vela pulsar.}
\citep[][]{Ackermann2010}. A typical duration of flares, which range
from a few days to several weeks, indicates that the spatial region
where a flare takes place is less than 1 pc (i.e., $c\times\Delta
t$, where $c$ is the speed of light and $\Delta t$ is a duration of
a flare). During episodes of strong \gray{} flares, AGNs are true
point sources to the \fermilat{} and the fluxes of the first
component strongly exceed the contribution of the second and third
components. Below, we study the \gray{} AGN emission during flares
and quiescent states to reconstruct the \fermilat{} PSF and to
perform a search for a pair halo component which might contribute to
quiescent AGN emission.

\section{Observations with Fermi-LAT and data reduction}

The Fermi satellite was launched on 2008 June 11 into a nearly
circular Earth orbit with an altitude of 565 km, an inclination
of 25.6$^{\circ}$, and an orbital period of 96 minutes. The
principal instrument on Fermi is the LAT \citep{Atwood2009}, a
pair-production telescope with a large effective area ($\sim$8000
cm$^2$ at 1 GeV) and field of view (2.4 sr). The energy range of LAT
sensitivity spans from 20 MeV to $>$300 GeV with an angular
resolution per single event of approximately $\approx5^{\circ}$ at
100 MeV and narrowing to $\approx0^{\circ}.8$ at 1 GeV, and
to $\approx0^{\circ}.15$ at 10 GeV. The LAT tracker has 12 layers
of 3\% radiation length tungsten converters (front section),
followed by 4 layers of 18\% radiation length tungsten converters (back section).
The front and back sections have intrinsically different PSF due to multiple
scattering with the PSF for front-converting events being approximately a factor of
two better than the PSF for back-converting events (e.g.,  an angular
resolution per single front-converting event of approximately $\approx0.6^{\circ}$ at
1 GeV and per single back-converting event of approximately $\approx1^{\circ}$ at
1 GeV). After the commissioning phase, the Fermi-LAT began routine science operations
on 2008 August 4. The Fermi-LAT normally operates in sky-survey mode
which provides a full-sky coverage every 3 hours (i.e., 2 orbits).

We downloaded the Pass 7 Reprocessed Fermi-LAT data from the Fermi
Science Support Center\footnote{\burl{http://heasarc.gsfc.nasa.gov/FTP/fermi/data/lat/weekly/photon/}}.
Note that the Pass 7 Reprocessed Fermi dataset uses updated
calibration constants \citep[][]{Bregeon2013}. The primary
differences with respect to the Pass 7 data are the correction of a
slight (1\% per year) expected degradation in the Calorimeter light
yield and significant improvement of the Calorimeter position
reconstruction, which in turn significantly improves the LAT
point-spead function at high energies ($>$5 GeV)
\footnote{\burl{http://www.slac.stanford.edu/exp/glast/groups/canda/lat_Performance.htm}}.
For the data analysis, we use the Fermi Science Tools v9r32p5
package\footnote{\burl{http://fermi.gsfc.nasa.gov/ssc/data/analysis/software/v9r32p5.html}}
and P7REP instrument response functions (IRFs). The improved
position reconstruction of \gray{} events above 5 GeV is
useful in the search for pair halos using the Pass 7 Reprocessed
Fermi-LAT data at these energies.

We used the Fermi LAT 2-year point source catalog
\citep[][]{Nolan2012}, gll\_psc\_v08.fit, to select the sample of
AGNs for our analysis. Firstly, we recorded the source type of all
2FGL sources and selected sources associated with AGNs. To reduce
the contamination by the Galactic diffuse foreground emission to the
regions of interest, we excluded AGNs with Galactic latitude
between -30$^{\circ}$ to 30$^{\circ}$. To reduce the contamination
the regions of interest by 2FGL \gray{} sources, we excluded AGNs
which have neighboring (2FGL) sources within 2$^{\circ}$. At this
stage, our sample contains 394 \gray{} AGNs. We began by selecting
all gamma rays of energy above 1 GeV within a 4$^{\circ}$ radius
around the direction of each AGN (the positions of AGNs taken from
the 2FGL catalog \citep[][]{Nolan2012}), and satisfying the SOURCE
event class. For this analysis, we have accumulated events obtained
from 2008 August 4 to 2013 November 7 (i.e., 5.2 years of the
\fermilat{} data), corresponding to 239557417 and 405478386 in units
of the Mission Elapsed Time. To reduce the contamination by the
\gray{} emission coming from cosmic-ray interactions in the Earth's
upper atmosphere (so-called albedo \grays{}) our selection is
refined by choosing events with zenith angles $<$100$^{\circ}$. We
removed events that occur during satellite maneuvers when the LAT
rocking angle was larger than 52$^{\circ}$. Time intervals when some
event has negatively affected the quality of the LAT data are also
excluded. Both the front-converting and back-converting events are
selected. Note that the statistics are higher when we consider both
the front-converting and back-converting events together. To justify
the usage of the front-converting and back-converting events together for
the increase of statistics, below we will compute the ratio of the numbers of
front-to-back events recorded during quiescent AGN states and recorded during
AGN flares, respectively, and will compare these ratios to make sure
that they are consistent within the error bars.
Below we show that the accumulation of the \fermilat{} data during
5.2 years made possible the reconstruction of the PSF (see Sect.
4.2) and the search for potential pair-halo extended emission
(see Sect. 3.2) by applying the technique based on a study of
quiescent and flaring AGN states.

\section{Temporal analysis for a search for pair halos around AGNs}

To disentangle AGN flares from quiescent AGN emission states, we
perform a temporal analysis of the \fermilat{} data. Below, we will
compare the spatial distribution of photons coming during AGN flare
episodes with that of photons coming during quiescent AGN states.
This will allow us to set the 95\% upper limits on the fraction of
\grays{} attributable to a pair halo component. To perform this
analysis, we will determine \gray{} flares for each AGN from our
sample of 394 AGNs and calculate the distribution of photons into
annular bins for AGN flares and for quiescent AGN states.

To disentangle AGN flares from quiescent AGN emission states, we
define an AGN flare in a statistical way. We determine flares in the
Poisson regime. The Poisson distribution expresses the probability
of a given number of events occurring in a fixed interval of
exposure assuming these events occur with the same average rate.
During AGN flare episodes, the flux from an AGN increases compared
with its quiescent state. To improve the strength of the selection
of flares, we calculate the total number of photons coming within
the circle with a radius of 1$^{\circ}$ above 1 GeV\footnote{We
use the energy threshold of 1 GeV for the selection of flares to
increase photon statistics. After the selection process of flares,
we will employ tight energy bands at higher energies to avoid the
systematic uncertainty caused by the presence of AGNs with different
spectra and by the steep energy dependence of the PSF below
$\simeq$5 GeV.}. For maximum flare-detection sensitivity, we select
only events within a relatively small spatial region of 1$^{\circ}$,
this is consistent with the selection choice of events for a
pulse-detection sensitivity
maximization\footnote{\burl{http://fermi.gsfc.nasa.gov/ssc/data/analysis/scitools/pulsar\_analysis_tutorial.html\#extractData}}.
Note that the 68\% containment angle for events of 1 GeV is about
$0.8^{\circ}$ for the Reprocessed Fermi-LAT data, while the 95\%
containment angle is about $2^{\circ}$. Note that the selection of
AGNs which have no neighboring 2FGL sources within 2$^{\circ}$ for
the study allows us to significantly suppress the effect of
neighboring sources on the detection of AGN flares, since less than
5\% of photons above 1 GeV from a neighboring source at a distance
of 2$^{\circ}$ can contribute to the region with a radius of
$1^{\circ}$ around a source from our sample.

To perform a search for AGN flares, we began by binning events in
equal exposure intervals which contain an average number of 5
photons. Note that the equal exposure intervals do not directly
correspond to equal time
intervals\footnote{\burl{http://fermi.gsfc.nasa.gov/ssc/data/analysis/LAT_caveats\_temporal.html}}
and, therefore, we computed equal exposure intervals. Our choice of
the equal exposure intervals containing 5 photons on average allows
us to study AGN flares in the Poisson regime. Using the Fermi
Science Tools, we calculated the exposure for each of the 394 AGNs.
The calculated exposure values are in the range of
$(155-229)\times10^9$ cm$^2$s. The mean exposure per AGN is
$171\times10^9$ cm$^2$s and the standard deviation is $15\times10^9$
cm$^2$s. The exposure intervals containing 5 photons are in the
range of $(0.4-3.0)\times10^8$ cm$^2$s with a mean of
$2.0\times10^8$ cm$^2$s and standard deviation of $0.5\times10^8$
cm$^2$s. The mean exposure interval roughly corresponds to 2 days
of \fermilat{} observations.

To define an AGN flare in a statistical way, we calculate
the probability of an observed number of photons in each equal
exposure interval from the Poisson distribution with the mean
number of photons. Furthermore, we multiply each of these
probabilities by a total number of equal exposure intervals
and then by 22, and compare the computed quantities with a value of 1.
If the computed quantity is less than 1, we record a flare occurrence
in the corresponding exposure interval.
Note that the multiplications performed here guarantee
that only $\approx$5\% of identical sources with the same average
number of photons contain such an excess of photons within one of
the exposure intervals. Note that the threshold value of
$\approx$5\% is chosen to guarantee that most of the computed
flaring intervals corresponds to the rate of photons that rarely
occurs in the Poisson regime and to make our selection of flaring
intervals sufficiently clean. When an AGN flare is detected, we
remove this exposure interval from the data set, re-calculate the
average number of photons, and repeat the procedure described above
to search for the next AGN flare. Finally, we record the start and
end times for all the detected flares for each AGN from our sample.
The total number of the computed flaring intervals is 965, only
$\approx 2\%$ of which are expected to be due to statistical
deviations in the Poisson process.

For the analysis of spatial distributions of photons for flares and
for quiescent states, we bin photons in three energy bands, namely,
4.5-6 GeV, 6-10 GeV, and $>$10 GeV. The selection of the highest
energy band, $E>10$ GeV, is motivated by the fact that multiple
scattering \citep[see][]{Atwood2009} is unimportant above 10 GeV and
the accuracy of the directional reconstruction of photon events
detected by \fermilat{} is limited by the ratio of the silicon-strip
pitch to silicon-layer spacing, whereas the first two energy bands
are selected as tight as possible to reduce the systematic error
caused by differences in the spectra of AGNs, yet still have enough
photons for a meaningful statistical analysis.
The 95\% containment angles for photon events from a point-like source
belonging to these energy bands are about 0.5$^{\circ}$\footnote{\burl{http://www.slac.stanford.edu/exp/glast/groups/canda/archive/p7rep_v15/lat_Performance_files/cPsfEnergy_P7REP_SOURCE_V15.pdf}}.
For each energy band, we bin photons in annular bins with radii of
$0^{\circ}<$r$<0^{\circ}.21$, $0^{\circ}.21<$r$<0^{\circ}.30$,
$0^{\circ}.3<$r$<0^{\circ}.42$, $0^{\circ}.42<$r$<0^{\circ}.52$,
$0^{\circ}.52<$r$<0^{\circ}.60$, and $1^{\circ}.0<$r$<1^{\circ}.3$.
Note the first two annular bins have a surface area twice smaller
than those of the 3rd, 4th, and 5th bins\footnote{the surface areas of
the 1st and 2nd annuli are chosen to be equal and the surface ares of the 3rd,
4th, and 5th annuli are chosen to be equal. For the sake of brevity,
in the paper we write the approximate values of the bin angular size.}.
This selection is motivated by the fact that the statistics for the first two bins is
sufficient to perform the analysis. We do not divide the 1st annular
bin into tighter annular bins in order to eliminate the effect
caused by the uncertainties in the determination of AGN positions on
the \gray{} sky in the 2FGL catalog. The outer annular bin is
selected for an estimation of the rate of background photons and has
a surface area which is $\approx$7.7 times larger than that of the
3rd, 4th, and 5th bins.

In Sect. 3.2, we demonstrate an application of the temporal analysis
based on the separation of photon events recorded during AGN flares
from those during quiescent states for a search for pair halos. Pair
halos around AGNs are expected to possibly reveal themselves as an
extended emission component in addition to a point source emission
from AGNs. Note that the size of the extended component should be
redshift-dependent. This dependence on a source redshift is caused
by 1) a redshift-dependent deflection angle of secondary electrons
and positrons in extragalactic magnetic fields (see Eqs. 30 and 31
from  \citet[][]{Neronov2009}), and 2) the geometry of propagation
of cascade \grays{} from the source to the observer (see Eq. 33 from
\citet[][]{Neronov2009}). Note that the motion of the secondary
electron-positron pairs is determined by the value of the
correlation length of the extragalactic magnetic field. Thus, if the
correlation length, $\lambda_{\mathrm{B}}$, is much less than
electron cooling distance $D_{\mathrm{e}}$, within which secondary
electrons/positrons lose their energy via IC scattering of the CMB
photons, deflections of secondary electrons/positrons are described
as diffusion in angle. In the opposite case, motion of secondary
electrons and positrons at the length scale of $D_{\mathrm{e}}$ can
be approximated by their motion in a homogeneous magnetic field. The
redshift dependencies are different for both these cases of motion.
Apart from the redshift dependence due to motion of secondary
electrons and positrons, geometry of \gray{} propagation introduces
another redshift dependent factor which is determined by the ratio
of a mean free path of primary TeV \grays{} propagating through the
EBL to a distance from the source. Below we perform two different
analyses to search for a pair halo component to compromise between
the number of detected photon events and the tightness of redshift
bins.

In Sect. 3.1, we start with a study of the spatial distributions of
photons recorded during AGN flare and quiescent states. In Sect.
3.2, we will bin the AGNs from our sample in several redshift bands
to search for pair halos in each of these selected
redshift-dependent samples. Finally, in Sect. 4.1, we will perform a
search for pair halos for individual strong AGNs included in our
sample.

\subsection{AGN flare and quiescent states}

To extract the approximate PSF on the basis of photons recorded
during AGN flares, we only use photons from AGNs that show the presence
of flaring activity. Our final sample then contains 158 AGNs including 56 BL Lac type blazars
of which 5 are definite identifications, 92 flat spectrum radio
quasars of which 6 are definite identifications, 7 active galaxies
of uncertain type, and 3 non-blazar AGNs of which 1 is a definite
identification. The ten brightest AGNs from our final sample,
corresponding to the 2FGL catalog, are PKS 1510-089, PKS 0537-441,
4C +21.35, Markarian 421, 3C 279, AO 0235+164, B2 1520+31, 3C 273,
PKS 1424+240, and PKS 0447-439.
\fermilat{} observations revealed the flaring activity of PKS
1510-089 \citep[][]{Abdo2010a}, 4C +21.35
\citep[][]{Ackermann2014a}, 3C 279 \citep[][]{ATel3C279}, B2 1520+31
\citep[][]{ATelB2, AtelB2a}, 3C 273 \citep[][]{Abdo2010b}, and PKS
1424+240 \citep[][]{Atel1424, Atel1424a}.

In this Section, we obtain the spatial distribution of \grays{}
recorded during AGN flares. This distribution should approximate
the PSF well (see Sect. 4.2). We also compare this spatial
distribution with that obtained from observations of a large sample
of AGNs during their quiescent states. A large sample of AGNs
contain \gray{} sources at different redshifts (see Sect. 3.2).
Since the size of pair halos strongly depends on redshift, the
contribution of pair halos with very different sizes to \gray{}
emission of a particular size is expected to be small. Below the
comparison between these distributions will establish that these
spatial distributions are similar, and further, we will proceed by
sorting AGNs in redshift bins in Sect. 3.2 to maximise the relative
contribution from pair halos to the \gray{} signal.

We stacked the photon events separately for flares and for quiescent
states for each selected energy band and each selected annular bin,
using our sample of the selected AGNs and the computed flaring and
quiescent states. The distribution of photons coming from AGN
flaring states is computed for our sample of 158 AGNs showing
flaring activity, while the distribution of photons coming from AGN
quiescent states is computed for the sample of the 394 selected AGNs
(see Sect.3). The stacked distributions of photons in annular bins
for flares and for quiescent states are presented in Table
\ref{Tab2}.

We used the stacked distributions of photons in annular bins for
flares and for quiescent states to derive the ratio of the number of
photons coming from AGNs and detected during quiescent states to the
number of photons coming from AGNs and detected during flares,
$R_{\mathrm{Q}/\mathrm{F}}$. We performed this procedure for each
annular bin and each selected energy band separately. The
statistical error taken was the square root of the count number. To
subtract background photons, we used the number of photons within
the outer annular bin, $1^{\circ}<r<1.3^{\circ}$, which is $\approx 7.7$
($\approx 15.4$) times larger than the 3rd, 4th, and 5th annular bins (the 1st and
2nd annular bins), respectively.
The errors in the determination of background
photon numbers were taken into account. The calculated numbers of
photons in annular bins for flares and for quiescent states for each
selected  energy band after subtraction of a background are shown in
Table \ref{Tab3}. Note that the calculated number of photons for AGN
flares provides us with similar statistics as provided by the Crab
pulsar and accumulated by \fermilat{} during 3.5 years (see Sect.
5.1). The calculated ratios, $R_{\mathrm{Q}/\mathrm{F}}$, and
statistical errors are shown in Table \ref{Tab4}. The calculated
ratios are within the error bars for each energy bin. We also
calculated the weighted mean of these ratios in the annular bin of
$0^{\circ}.21<r<0^{\circ}.6$ for each energy band and the variance
of the mean. The weighted mean of the ratios agrees with the ratio
computed for the first annular bin (this bin contains the
highest number of photons amongst all the bins, and therefore,
provides the smallest statistical error).

Note that the ratio of the number of photons from AGNs to that from
a background is much higher for flares than for quiescent states and
that the ratio of the number of photons from AGNs to that from a
pair halo component is much higher for flares than for quiescent
states. These conclusions about the ratios are equivalent to the
suggestion that the flux from an AGN (from our sample) during
its flaring state is higher than that during its quiescent state.
We found that 34 bright flaring AGNs from our sample during their
flaring states gives 80\% of \grays{} above 4.5 GeV recorded
during the flaring states of all 158 AGNs in the sample. To establish that the average flux
during the flaring state of an AGN is higher than that during the quiescent
state, we computed the ratio of the average flux recorded during
flares to that during the quiescent state for each of these 34
bright flaring AGNs. We found that these ratios exceed 5 for all
these 34 AGNs, while the ratios exceed 10 for 18 of these 34 AGNs.
The ratio of the mean flux recorded during  flaring states to that recorded
during quiescent states for other flaring AGNs of our sample is
$\approx 5$ \footnote{Note that this does not mean that the factor of 5 difference
in ``flaring" versus ``quiescent" states is a result of the selection of the
mean number of photons of 5 in each exposure interval, i.e. the mean flux ratio between flaring and quiescent states would not become 10, if the mean number of photons of 10 in each exposure interval is selected. Thus, for example if a flare is short and lies within one exposure interval then the flaring-to-quiescent flux ratio will be higher for a smaller mean number of photons within one exposure interval. On the other hand, if a flare is long, for example with the duration half a total exposure then the flaring-to-quiescent flux ratio will not depend on the mean number of photons within one exposure interval.} The fact that it is more promising to search for a
pair halo during quiescent states than to search for a pair halo
during flaring states was first emphasized by \citet{Aharonian1994} and was
used by \citet{Barbiellini2014} in their analysis of 3C 279.

We also noted that the ratio of the number of photons coming from
AGNs and detected during quiescent states to that detected during
flares is higher for the energy band of $E>10$ GeV compared with
those for the energy bands of 4.5 GeV$<E<$6 GeV and 6 GeV$<E<$10
GeV. A probable explanation is the presence of the selection bias
because the sample of 394 sources contains 185 BL Lac type blazars
and 146 FSRQs, while the sample of 158 sources contains 56 BL Lac
type blazars and 92 FSRQs.

We also repeated our analysis and computed the ratio of the number
of photons coming from AGNs and detected during quiescent states to
the number of photons coming from AGNs and detected during flares by
using only the sample of the 158 AGNs showing flaring activity
(this guarantees the same ratio of the numbers of BL Lac type
blazars to FSRQs for which flares and quiescent states are recorded).
This is useful to suppress a possible selection effect due to
considering different AGN samples for flaring and quiescent states.
These ratios are shown in Table \ref{TabNEW}. The calculated ratios
are also within the error bars for each energy bin. The ratios shown
for the first annular bin are statistically significantly different
between the energy band, $>10$GeV, and the other two energy bands.
However, a detailed study of this question is beyond of the scope of
our paper. Note that the ratio computed for the numbers of
background photons (i.e. those from the outer annuli) recorded
during AGN flares and quiescent states does reveal any change
from one energy band to another. This indicates that the
changing of the ratio with energy as described is related to the AGN themselves
rather than to any background variation in time.

\begin{table}
\centering \caption{The ratio of the number of photons coming from
158 AGNs and detected during quiescent states to the number of photons
coming from AGNs and detected during flares}
\begin{tabular}{ | c | c | c | c |}
\hline
& \multicolumn{2}{r}{Energy bands} \\
\cline{2-4}
Annular bin & $4.5-6$ GeV & $6-10$ GeV & $>10$ GeV\\
\hline
$0^{\circ}<$r$<0^{\circ}.21$ & $3.2\pm0.2$ & $3.38\pm0.15$ & $4.7\pm0.2$ \\
$0^{\circ}.21<$r$<0^{\circ}.3$ & $2.9\pm0.3$ & $2.9\pm0.3$ & $5.2\pm0.7$ \\
$0^{\circ}.3<$r$<0^{\circ}.42$ & $2.8\pm0.3$ & $3.7\pm0.5$ & $4.9\pm0.9$ \\
$0^{\circ}.42<$r$<0^{\circ}.52$ & $4.3\pm0.9$ & $3.1\pm0.7$ & $4.7\pm1.1$ \\
$0^{\circ}.52<$r$<0^{\circ}.6$ & $3.3\pm1.0$ & $4.5\pm1.4$ & $2.5\pm1.3$ \\
\hline
\end{tabular}
\label{TabNEW}
\end{table}

\subsection{Search for pair halos around AGNs within several redshift intervals}

The sizes of pair halos depend on the source redshifts and,
therefore, to search for pair halos around AGNs the redshifts should
be taken into account. Below, we present the temporal analysis for
samples of AGNs with known redshifts belonging to several redshift
intervals. Using the second catalog of AGNs detected by \fermilat{}
\citep[][]{AgnCatalog} we select only the AGNs with known redshifts
from our sample of 394 AGNs. Since the number of photon events
coming during AGN flares is much smaller ($\sim5$ times) than that
that coming during quiescent states (see Table \ref{Tab3}), we will
use all 158 AGNs showing \gray{} flaring activity to approximate the
PSF using flares of AGNs. To have a higher number of photons during
AGN quiescent states, we select all AGNs with known redshifts from
our sample of 394 AGNs. We sort AGNs in redshift intervals only for
quiescent states.

To perform the analysis within several redshift intervals, the
redshift intervals should be as tight as possible to expect a
similar angular extension of pair-halos within each redshift
interval and should be as broad as possible to increase the number
of photon events from AGNs within similar redshifts. In this paper,
we sort the AGNs with known redshifts belonging to our sample into
four redshift intervals, $z<0.2$, $0.2<z<0.6$, $0.6<z<1.3$, and
$1.3<z$. The numbers of AGNs in these redshift intervals are 34, 60,
83, and 57, respectively, i.e., 234 AGNs with known redshifts in total.
Note that these redshift intervals are
tighter than those chosen by \citet[][]{Ando2010} which were $z<0.5$
and $0.5<z$. We introduce two redshift intervals to gather nearby
AGNs $(z<0.2)$ in one of these two intervals and very distant AGNs
$(z>1.3)$ in the other interval. Since the size of extended
emission from pair halos steeply decreases with distance and is
proportional to the strength of extragalactic magnetic field
\citep[see, e.g.,][]{Neronov2009}, the introduction of these two
redshift intervals allows us to cover the cases of a lower and a
higher strength of extragalactic magnetic field. We estimated the
strength of extragalactic magnetic field which can be tested using
the present analysis by assuming that the angular extension of pair
halos within each redshift interval is $0.3^{\circ}$ and that the
correlation length, $\lambda_{\mathrm{B}}$, is much less than the
electron cooling distance $D_{\mathrm{e}}$. Putting the values of
the redshift and angular extension in Eq. 35 of
\citep[][]{Neronov2009}, we found that the values of strength of
extragalactic magnetic field which can be tested by using this
analysis,
$B\times\left(\lambda_{\mathrm0}/1\mathrm{kpc}\right)^{1/2}$, are
about $10^{-15}$, $10^{-14}$, $5\times10^{-14}$, and $10^{-13}$
gauss for the redshift intervals of $z<0.2$, $0.2<z<0.6$,
$0.6<z<1.3$, and $1.3<z$, respectively\footnote{To obtain the
magnetic field strength which can be tested by using AGNs in the
redshift interval of $z<0.2$, we assume that the strongest sources,
Mkn 421 and Mkn 501, give the strongest contribution to pair halo
emission.}. The stacked distributions of photons in annular bins for
quiescent states of AGNs within the four redshift intervals are
presented in Table \ref{TabStackRedshift}.

To justify the usage of the front-converting and back-converting
events together for the increase of statistics, we compute the ratio
of the number of front-converting events to that of back-converting
events (the front-to-back ratio) recorded during quiescent AGN states
for each redhift interval and recorded during AGN flares, respectively.
We compute the front-to-back ratios for each energy band individually
and compare these ratios computed for quiescent states and flares.
The front-to-back ratios for flares are $1.50\pm0.10$ for the 4.5-6 GeV band,
$1.42\pm0.09$ for the 6-10 GeV band, and $1.54\pm0.11$ for the $>10$ GeV band,
while the central values of the front-to-back ratios for quiescent states
are in the range of $\approx1.30-1.45$ for each of the redshift intervals and each
of the energy bands. We checked that the front-to-back ratios for quiescent states
lie within the 2$\sigma$ error bars around the central values for the front-to-back
ratio computed for flares (i.e., are consistent with the 2$\sigma$ confidence level).
Since the PSF for front-converting events are tighter than that for back-converting
events, the effect of an excess of the front-to-back ratio for flares compared with that
for quiescent states could mimic the presence of a pair halo. We do not
detect this effect (neither do we detect pair halos) using the present data
and conclude that this effect is subdominant.
Although this effect does not produce an observable spurious extension of the sources
during their quiescent states, this effect can still produce a systematic
bias on the upper limit on pair halos derived in the paper. Since the existing data
does not show any significant difference in the front-to-back ratio
for flares and for quiescent states, we leave this question for studying in the future
when higher statistics will be accumulated.

To compare the spatial distributions of photons detected during
quiescent states of AGNs within the given redshift interval with
that of photons detected during the flares of AGNs from our final
sample of 158 AGNs, we divide these numbers by each other after
subtraction of a background. The ratios of these numbers of photons
for each annular bin and each energy band are shown in Table
\ref{RatioRedshift}. Note that background rates and statistical
errors were estimated in the same way as the analysis of photons
from AGNs belonging to our final sample presented above. If the
relative error on the ratio is high, we do not list such a value in
this Table. The calculated ratios are in general within the error
bars for each energy bin and each redshift interval. Therefore, no
evidence for a pair halo component is found in any of the selected
redshift intervals.

\begin{table*}
\centering \caption{The ratio of the number of photons detected
during quiescent states of AGNs within the given redshift interval
with that of photons detected during the flares of AGNs from our
entire sample.}
\begin{tabular}{ | c | c | c | c | c |}
\hline
& \multicolumn{3}{r}{Energy bands} \\
\cline{3-5}
Redshift & Annular bin & $4.5-6$ GeV & $6-10$ GeV & $>10$ GeV\\
\hline
$z<0.2$ & $0^{\circ}<$r$<0^{\circ}.21$ & $0.85\pm0.05$ & $1.07\pm0.06$ & $2.15\pm0.11$ \\
$z<0.2$ & $0^{\circ}.21<$r$<0^{\circ}.3$ & $0.82\pm0.09$ & $0.89\pm0.11$ & $2.28\pm0.35$ \\
$z<0.2$ & $0^{\circ}.3<$r$<0^{\circ}.42$ & $0.66\pm0.10$ & $0.94\pm0.15$ & $2.35\pm0.43$ \\
$z<0.2$ & $0^{\circ}.42<$r$<0^{\circ}.52$ & $0.99\pm0.27$ & $1.03\pm0.28$ & $2.03\pm0.53$ \\
$z<0.2$ & $0^{\circ}.52<$r$<0^{\circ}.6$ & $-$ & $-$ & $1.02\pm0.65$ \\
\hline
$0.2<z<0.6$ & $0^{\circ}<$r$<0^{\circ}.21$ & $0.90\pm0.06$ & $0.93\pm0.05$ & $1.36\pm0.07$ \\
$0.2<z<0.6$ & $0^{\circ}.21<$r$<0^{\circ}.3$ & $0.81\pm0.13$ & $0.74\pm0.09$ & $1.44\pm0.22$ \\
$0.2<z<0.6$ & $0^{\circ}.3<$r$<0^{\circ}.42$ & $0.75\pm0.10$ & $1.16\pm0.16$ & $1.29\pm0.24$ \\
$0.2<z<0.6$ & $0^{\circ}.42<$r$<0^{\circ}.52$ & $1.51\pm0.30$ & $1.23\pm0.25$ & $1.53\pm0.36$ \\
$0.2<z<0.6$ & $0^{\circ}.52<$r$<0^{\circ}.6$ & $0.83\pm0.25$ & $2.10\pm0.57$ & $2.14\pm0.70$ \\
\hline
$0.6<z<1.3$ & $0^{\circ}<$r$<0^{\circ}.21$ & $1.04\pm0.06$ & $1.07\pm0.06$ & $1.17\pm0.07$ \\
$0.6<z<1.3$ & $0^{\circ}.21<$r$<0^{\circ}.3$ & $0.95\pm0.11$ & $1.11\pm0.13$ & $1.30\pm0.23$ \\
$0.6<z<1.3$ & $0^{\circ}.3<$r$<0^{\circ}.42$ & $0.90\pm0.15$ & $1.02\pm0.20$ & $1.26\pm0.31$ \\
$0.6<z<1.3$ & $0^{\circ}.42<$r$<0^{\circ}.52$ & $1.39\pm0.40$ & $0.83\pm0.31$ & $1.08\pm0.43$ \\
$0.6<z<1.3$ & $0^{\circ}.52<$r$<0^{\circ}.6$ & $-$ & $1.50\pm0.72$ & $-$\\
\hline
$1.3<z$ & $0^{\circ}<$r$<0^{\circ}.21$ & $0.58\pm0.04$ & $0.41\pm0.03$ & $0.43\pm0.03$ \\
$1.3<z$ & $0^{\circ}.21<$r$<0^{\circ}.3$ & $0.51\pm0.07$ & $0.37\pm0.07$ & $0.31\pm0.11$ \\
$1.3<z$ & $0^{\circ}.3<$r$<0^{\circ}.42$ & $0.48\pm0.10$ & $0.58\pm0.13$ & $0.60\pm0.21$ \\
$1.3<z$ & $0^{\circ}.42<$r$<0^{\circ}.52$ & $0.90\pm0.29$ & $0.77\pm0.26$ & $-$ \\
$1.3<z$ & $0^{\circ}.52<$r$<0^{\circ}.6$ & $0.63\pm0.40$ & $0.97\pm0.29$ & $-$ \\
\hline
\end{tabular}
\label{RatioRedshift}
\end{table*}

\subsubsection{Upper limits for the fraction of $\gamma$ rays attributable to
a pair halo component}

To set the upper limits for the fraction of \grays{} attributable to
a pair halo component, we should make an assumption about the
spatial profile of a pair halo \citep[e.g., see][]{Ackermann2013}.
Note that if the pair halo profile is narrow compared with the PSF,
it would be difficult to disentangle a point-like component from a
pair halo. Therefore, in this analysis we assume that a pair halo is
an extended component to \fermilat{} with an angular size of
$\gtrsim0.3^{\circ}$ (which exceeds the radius of the second annular
bin). Since the signal-to-noise ratios of the \gray{} signal from
the stacked source centered on AGNs are highest for the first and
second annular bins, we will use these two annular bins to set the
upper limits for the fraction of \grays{} attributable to a pair
halo component. Note that the fluxes from AGNs during episodes of
\gray{} flares are much higher than those during quiescent states
due to the definition of flares (see the beginning of Section 3)
and, therefore, the contribution of a pair halo component to flare
fluxes is negligible compared to that of quiescent state fluxes.
Thus, we assume that the contribution to the first and second
annular bins from a pair halo component are the same for quiescent
AGN states and that the contribution from a pair halo component to
AGN flares is negligible. First we divided the number of photons
after the subtraction of a background in the first bin by that in
the second bin for AGN flares and for quiescent AGN states. Second
we subtracted the equal number of photons, $x$, from the first and
second annular bins for quiescent AGN states (these photons come
from a possible pair halo component). After this procedure, the
ratios of the number of photons must be the same. Thus, this
procedure leads to the equation for the ratios of the numbers of
photons in the first and second annular bins for the energy band of
4.5-6 GeV and for the redshift interval, z$<$0.2,
\begin{equation}
\frac{504-19\pm22-x}{165-19\pm13-x}=\frac{570\pm24}{178\pm14},
\label{eqhalo}
\end{equation}
where the left-hand side of this equation is for quiescent states
(see Table \ref{TabStackRedshift}) and the right-hand side is for
flares (see Table \ref{Tab3}). The equations for the energy band of
6-10 GeV and $>$10 GeV were obtained using the same procedure.
Taking the error bars into account, we computed the 95\% upper limit
for the fractions of \grays{} from the stacked AGN sample
attributable to a pair halo component in the first annular bin
$(r<0.21^{\circ})$ for each selected redshift interval.
The computed 95\% upper limits are shown in the upper part of Table \ref{RatioHaloes}.

\begin{table*}
\centering \caption{The 95\% upper limits for the fractions of
\grays{} from the stacked AGN sample attributable to a pair halo
component in the first annular bin $(r<0.21^{\circ})$ for each
selected redshift interval.}
\begin{tabular}{ | c | c | c | c | c |}
\hline
& \multicolumn{3}{r}{Energy bands} \\
\cline{3-5}
Redshift & background & $4.5-6$ GeV & $6-10$ GeV & $>10$ GeV\\
\hline
$z<0.2$ & subtracted & $12\%$ & $7\%$ & $6\%$ \\
$0.2<z<0.6$ & subtracted & $11\%$ & $7\%$ & $10\%$ \\
$0.6<z<1.3$ & subtracted & $11\%$ & $9\%$ & $6\%$ \\
$1.3<z$ & subtracted & $15\%$ & $11\%$ & $7\%$ \\
\hline
$z<0.2$ & included & $13\%$ & $7\%$ & $6\%$ \\
$0.2<z<0.6$ & included & $11\%$ & $8\%$ & $10\%$ \\
$0.6<z<1.3$ & included & $15\%$ & $12\%$ & $12\%$ \\
$1.3<z$ & included & $20\%$ & $17\%$ & $15\%$ \\
\hline
\end{tabular}
\label{RatioHaloes}
\end{table*}

Note that the obtained constraints on the fraction of \grays{}
attributable to a pair halo component were derived under the
assumption that a pair halo component does not contribute to the
region of $1^{\circ}.0<$r$<1^{\circ}.3$. This region was used to
estimate a background. Using the number of photons in the outer
annular bin (see Table \ref{TabStackRedshift}), we found that the
fraction of \grays{} from the stacked AGN sample attributable to a
background in the first annular bin ($r<0^{\circ}.21$) is 2-4\% for
the three energy bands and for the redshift interval of $z<0.2$,
4-6\% for the three energy bands and for the redshift interval of
$0.2<z<0.6$, 6-7\% for the three energy bands and for the redshift
interval of $0.6<z<1.3$, and 9-11\% for the three energy bands and
for the redshift interval of $z>1.3$. The assumption that a pair
halo component does not contribute to the region of
$1^{\circ}.0<$r$<1^{\circ}.3$ can be weakened. For this purpose, we
modified the procedure described above. First we divided the number
of photons without the subtraction of a background (see Table
\ref{Tab3}) in the first bin by that in the second bin for AGN
flares and for quiescent AGN states. Second we subtracted the equal
number of photons, $x$, from the first and second annular bins for
quiescent AGN states (these photons come from a possible pair halo
component + a background). Note that a background contributes the
same number of photons in the first and second annular bins, because
their surface areas are the same. Thus, this procedure leads to the
equation for the ratios of the numbers of photons in the first and
second annular bins for the energy band of 4.5-6 GeV and for the
redshift interval, z$<$0.2,
\begin{equation}
\frac{504\pm22-x}{165\pm13-x}=\frac{574\pm24}{182\pm14},
\end{equation}
where the left-hand side of this equation is for quiescent states
(see Table \ref{TabStackRedshift}) and the right-hand side is for
flares (see Table \ref{Tab2}). We compute the 95\% upper limits for
the fraction of \grays{} from the stacked AGN sample attributable to
a pair halo component + a background in the first annular bin
($r<0^{\circ}.21$). The computed 95\% upper limits are shown in
the bottom part of Table \ref{RatioHaloes}.

\subsubsection{Validation with a likelihood analysis}

To strengthen the results obtained above, we use a likelihood
analysis. This analysis permits us to search for a pair halo
component around AGNs in the LAT telescope data and to set upper
limits on the fraction of photons attributable to this component.
Likelihood is the joint probability of the observed data given the
hypothesis and quantifies the relative extent to which the data
supports a statistical hypothesis. Below we consider the null
hypothesis which implements that gamma-ray emission from AGNs comes
from compact sources which are point-like gamma-ray sources for the
LAT telescope. As the alternative hypothesis, we consider that
gamma-ray emission from AGNs comes from both the compact sources and
extended pair halo component. A likelihood function (often the
likelihood) is a function of the parameters of a statistical model
(in our case, these parameters are the normalisations of a
point-like component and of a pair halo component).

We use the same 234 AGNs with known redshifts which were used
above in this Section. As done above, we sort these AGNs into four
redshift intervals, $z<0.2$, $0.2<z<0.6$, $0.6<z<1.3$, and $1.3<z$
and bin photons in three energy bands, namely, 4.5-6 GeV, 6-10 GeV,
and $>10$ GeV. To perform a likelihood analysis, we sort photons in
six annuli, $0<r<0.05^{\circ}$, $0.05<r<0.1^{\circ}$,
$0.1<r<0.15^{\circ}$, $0.15<r<0.20^{\circ}$, $0.20<r<0.25^{\circ}$,
and $0.25<r<0.30^{\circ}$. We have performed maximum likelihood
analysis assuming that, in addition to the central point sources and
diffuse backgrounds, there is a pair halo component. We fit the
histogram of photon counts as a function of $\theta^2$ by
minimising
\begin{equation}
\chi^{2}=\sum_{\mathrm{i}} \frac{(N_{\mathrm{i}}-N_{\mathrm{model,
i}})^2}{N_{\mathrm{i}}},
\end{equation}
and
\begin{equation}
N_{\mathrm{model, i}}=N_{\mathrm{psf}}
P_{\mathrm{psf}}(\theta^2_{\mathrm{i}})+N_{\mathrm{halo}}
P_{\mathrm{halo}}(\theta^2_{\mathrm{i}})+N_{\mathrm{bg, i}},
\end{equation}
where $N_{\mathrm{psf}}$ and $N_{\mathrm{halo}}$ are free
parameters of the alternative model (see also \citet[][]{Ando2010}).
As for the null model $N_{\mathrm{psf}}$ is a free parameter and the parameter
$N_{\mathrm{halo}}$ equals zero by definition. The index $i$ refers
to the i-th annular bin, $N_{\mathrm{i}}$ is the total number of
events in the annular bin recorded during flaring states.
$P_{\mathrm{psf}}$ is the normalised PSF and $N_{\mathrm{bg, i}}$ is
the number of events due to diffuse backgrounds. To estimate the
rate of background photons, we use the outer (large) annular bin,
$1.0<r<1.3^{\circ}$, and compute the number of background photons in
each of six annular bins by using surface area scales. In this
analysis we assume that a pair halo, described as
$P_{\mathrm{halo}}$, is an extended component to Fermi-LAT with an
angular size of $\gtrsim 0.3^{\circ}$ and its contribution to each
of the six annular bins is proportional to the surface areas of
these annular bins, respectively. The normalised PSF is taken from
the distribution of photons recorded during AGN flares.

From the result of the $\chi^{2}$ minimisation, we found that
the minimised $\chi^{2}$ values agree with the expected values, i.e.
the computed $\chi^{2}$ are typically in the range of
($\mathrm{d.o.f.}-\sqrt{2 \mathrm{d.o.f.}}$,
$\mathrm{d.o.f.}+\sqrt{2 \mathrm{d.o.f.}}$), where $\mathrm{d.o.f.}$
is the number of degrees of freedom. This means that the fits
describe the observed data rather well. The only exception is with
$\chi^{2}\approx20$, which occurs for nearby AGNs, $z<0.2$, and for the highest
energy band, $E>10$ GeV. Note that there is a strong contribution of
the source, Mkn 421, in the 1st redshift interval at high energies,
$E>10$ GeV for quiescent states. Mkn 421 is a very hard spectrum
gamma-ray source with a photon index of $\approx1.77$ and its semiminor
and semimajor axes at 68\% confidence are of $0.0067^{\circ}$
as derived in \citet[][]{Nolan2012}. Semiminor and semimajor
axes of many 2FGL sources are derived with one order of magnitude
higher uncertainties than those for Mkn 421 in the 2FGL catalog
\citet[][]{Nolan2012}. We noted that the discrepancy between the
observation and model is particularly strong in the annular bin,
$r<0.05^{\circ}$, for the redshift interval $z<0.2$ and for the
highest energy band, $E>10$ GeV. If we exclude photons from Mkn 421,
then the minimised $\chi^{2}$ value is 7.5 and is consistent with
the expected one. In the limit of a large number of counts in each
bin, the likelihood is given by $\mathcal{L}=exp(-\chi^{2}/2)$, so
that minimising $\chi^{2}$ is equivalent to maximising the
likelihood, $\mathcal{L}$. We found that the inclusion of a pair
halo component in the model does not improve the likelihood value
sufficiently to establish the presence of this pair halo component
in the data. Therefore, we derived the one-sided 95 per cent upper
limit on the fraction of photons attributable to a pair halo
component by fitting the normalisation of this component, for which
we increase its normalisation until the maximum likelihood decreases
by 2.71/2 in logarithm. The computed upper limits are between 2\%
and 6\% depending on energy band and redshift interval. These upper
limits are stronger than those obtained before. Note that the model
for a point-like source used in the likelihood analysis is
considered to be precisely established, however the number of
photons recorded during flaring states is close to those numbers of
photons recorded during quiescent states for each of these redshift
intervals. The expression, such as Eq.\ref{eqhalo}, leads to
more conservative upper limits on the fraction of photons
attributable to a pair halo component, since it takes the error bars
assigned to the model into account. If the point-like source model
is considered well established, then the error bars shown in Table
\ref{RatioHaloes} would decrease by a factor of $\approx$1.5).

We also checked if the computations presented above are
reproducable by maximising the Poisson log-likelihood function,
$l=\ln{\mathcal{L}}$, where
\begin{equation}
l=\sum_{\mathrm{i}} N_{\mathrm{i}} \ln{N_{\mathrm{model, i}}}-\sum_{\mathrm{i}}
N_{\mathrm{model, i}}.
\end{equation}
Maximising the Poisson log-likelihood function, we found that
the best-fit normalisations of the point-like component and the
upper limits on the pair halo component agree well with those
derived by using the $\chi^{2}$ minimisation.

\section{Method to reconstruct the Fermi-LAT PSF}

In this Section, we perform a temporal analysis of each of the five
brightest \gray{} AGNs from our sample in order to search for pair
halos around each of these sources. Further we reconstruct the
\fermilat{} PSF which is not affected by a pair halo component.

As was mentioned above, the contribution of a pair halo component to
the total AGN \gray{} emission is much smaller for flares than that
for quiescent states. Therefore, the spatial distribution of photons
coming during AGN flares can be used as a proxy of the PSF.

The spatial distribution of photons coming during quiescent AGN
states, $D_{\mathrm{Q}}(x)$, for each energy band and each source
can be written as
\begin{equation}
D_{\mathrm{Q}}(x)=\alpha_{\mathrm{Q}}
\mathrm{PSF}(x)+H_{\mathrm{Q}}(x)+B_{\mathrm{Q}}(x), \label{eqq}
\end{equation}
where the first, second, and third terms are the contributions of a
point source, a pair halo, and a background, respectively, and $x$
is the label for annular bins, $\alpha_{\mathrm{Q}}$ is the
normalisation of a point source flux. Note that the PSF component is
not affected by a pair halo component. Similarly the spatial
distribution of photons coming during AGN flares,
$D_{\mathrm{F}}(x)$, can be written as
\begin{equation}
D_{\mathrm{F}}(x)=\alpha_{\mathrm{F}}
\mathrm{PSF}(x)+H_{\mathrm{F}}(x)+B_{\mathrm{F}}(x). \label{eqf}
\end{equation}
The pair halo and background fluxes are time-invariant and, therefore,
\begin{equation}
\frac{H_{\mathrm{F}}(x)+B_{\mathrm{F}}(x)}{H_{\mathrm{Q}}(x)+B_{\mathrm{Q}}(x)}=\frac{N_{\mathrm{F}}}{N_{\mathrm{Q}}},
\end{equation}
where $N_{\mathrm{Q}}$ and $N_{\mathrm{F}}$ are the total number of
equal exposure intervals for quiescent states and for flares,
respectively, for the given AGN. Multiplying Eqs. \ref{eqq} and
\ref{eqf} by $N_{\mathrm{F}}$ and $N_{\mathrm{Q}}$, respectively,
and subtracting one equation from the other, we find
\begin{equation}
\mathrm{PSF}(x)=\frac{D_{\mathrm{F}}(x) N_{\mathrm{Q}} -
D_{\mathrm{Q}}(x) N_{\mathrm{F}}}{\alpha_{\mathrm{Q}} N_{\mathrm{F}}
- \alpha_{\mathrm{F}} N_{\mathrm{Q}}}. \label{eqpsf}
\end{equation}
Note that the ratio of the values of the point spread function for
two annular bins depends only on the values of ${D_{\mathrm{F}}(x)
N_{\mathrm{Q}} - D_{\mathrm{Q}}(x) N_{\mathrm{F}}}$ for these
annular bins. This fact can be used to reconstruct the PSF by
computing the ratios of the PSF values for different annular bins.

\subsection{Analysis of individual sources}

To outline the method of the PSF reconstruction, below we calculate
the ratio, $R$, of the values of the point spread function for the
two annular bins of $0<r<0^{\circ}.21$ and
$0^{\circ}.21<r<0^{\circ}.42$. Note that the later annular bin
corresponds to the sum of the second and third annular bins from
Table \ref{Tab3}. For this study, we selected the five brightest
sources from our final sample, namely 4C +21.35, 3C 279, PKS
1510-08, PKS 0537-441, and Mkn 421. We calculate the ratio, $R$, for
each of these sources and the weighted mean of these ratios. For an
illustration, we selected photons in the energy band of 4.5-6 GeV
and photons in the energy band of 6-10 GeV. The observed numbers of
photons in the two annular bins of $0<r<0^{\circ}.21$ and
$0^{\circ}.21<r<0^{\circ}.42$ for flares and for quiescent states
for the five brightest sources are shown in Table \ref{Tab5}.  Using
Eq. \ref{eqpsf} and Table \ref{Tab5}, we calculated the ratios of
the values of the point spread function for these two annular bins,
$R$, for the five brightest sources. The calculated ratios are shown
in Table \ref{Tab6}. The weighted mean of the ratios for these five
sources is $\bar{R}=1.7\pm0.3$ for the energy band of 4.5-6 GeV and
is $\bar{R}=2.7\pm0.4$ for the energy band of 6-10 GeV. We compared
the calculated weighted mean of the ratios for these five sources
for the energy band of 4.5-6 GeV, $\bar{R}=1.7\pm0.3$, with the
ratio of the numbers of photons in these two annular bins for flares
and for quiescent states (see Table \ref{Tab3}). The ratios of the
numbers of photons in these two annular bins coming during flares
and quiescent states are $R_{F}=1.9\pm0.1$ and $R_{Q}=2.2\pm0.1$,
respectively. All these three ratios ($\bar{R}$, $R_{F}$, and
$R_{Q}$) agree within the error bars. Note that the value of
$\bar{R}$ is close to that of $R_{\mathrm{F}}$ as was expected. We
also compared the calculated weighted mean of the ratios for these
five sources for the energy band of 6-10 GeV, $\bar{R}=2.7\pm0.4$,
with the ratio of the numbers of photons in these two annular bins
for flares and for quiescent states, $R_{F}=2.7\pm0.2$ and
$R_{Q}=2.8\pm0.1$, respectively. These three values ($\bar{R}$,
$R_{F}$, and $R_{Q}$) agree within the error bars for the energy
band of 6-10 GeV. Therefore, no evidence for a pair halo is found by
means of these comparisons.

We also obtained the constraints on the fraction of \grays{}
attributable to a pair halo component for each of the five brightest
AGNs. We divided the number of photons without the subtraction of a
background (see Table \ref{Tab5}) in the first bin by that in the
second bin for quiescent AGN states and for AGN flares. Then we
subtracted the number of photons, $x$, from the first annular bin
and subtracted the number of photons, $3x$, from the second annular
bin \footnote{The factor of 3 is due to a larger surface area of the
second annular bin compared with that of the first annular bin.} for
quiescent AGN states (these photons come from a possible pair halo
component + a background). To improve the constraints we used our
final sample of 158 AGNs to compute the number of photons during AGN
flares. We found that the 95\% upper limit for the fraction of
\grays{} from the stacked AGN sample attributable to a pair halo
component + a background in the first annular bin $(r<0.21^{\circ})$
equals 20\% for the energy band of 4.5-6 GeV and 7\% for the energy
band of 6-10 GeV for 4C+21.35. For the other four selected AGNs
these ratios for the energy bands of 4.5-6 GeV and 6-10 GeV are 10\%
and 19\% for 3C 279, 13\% and 7\% for PKS 1510-08, 7\% and 5\% for
PKS 0537-441, and 12\% and 4\% for Mkn 421. Note that as a proxy of
the PSF for the computation of the upper limits on the fraction of
\grays{} attributable to a pair halo component, we used the spatial
distribution of photons coming from AGNs during their flares.

\begin{table}
\centering \caption{The ratios, $R$, of the values of the point
spread function for the first two annular bins and for the energy
bands of 4.5-6 GeV and 6-10 GeV.}
\begin{tabular}{ | c | c | c | }
\hline
Source name & $R$ for 4.5-6 GeV & $R$ for 6-10 GeV\\
\hline
4C +21.35 & 1.5$\pm$0.4 & 2.9$\pm$0.8\\
3C 279 & 1.5$\pm$0.6 & 1.9$\pm$1.0\\
PKS 1510-08 & 2.7$\pm$0.8 & 2.4$\pm$0.7\\
PKS 0537-441 & 2.2$\pm$0.8 & 3.8$\pm$1.2\\
Mkn 421 & 2.6$\pm$1.5 & 4.4$\pm$2.1\\
\hline
weighted mean & 1.7$\pm$0.3 & 2.7$\pm$0.4\\
\hline
\end{tabular}
\label{Tab6}
\end{table}

\subsection{Reconstruction of the Fermi-LAT PSF}

The onboard data obtained by the LAT instrument can be used to
derive its instrument response functions, such as the PSF. To
perform a reconstruction of the PSF, one needs to select an
astrophysical \gray{} source which is point-like for the
instrument. One class of \gray{} sources used for the PSF
reconstruction is pulsar; another class used for this is AGN.
However, the presence of pair halos around AGNs limits the
application of AGNs for the PSF reconstruction. In this Section, we
demonstrate how to reconstruct the \fermilat{} PSF that is not
affected by a pair halo component by stacking 158 flaring AGNs of
our sample.

We take each of the 158 flaring sources of our sample and derive the
angular distribution of \grays{} for each source for both flaring
and quiescent states. Then we compute the value of
$D_{\mathrm{F}}(x) N_{\mathrm{Q}} - D_{\mathrm{Q}}(x)
N_{\mathrm{F}}$ for each source and for each annular bin. This gives
the ``pure'' PSF profile, free  of the possible extended pair-halo
signal. Then we stack the PSF profile obtained in this way on a
source-by-source basis for all the 158 sources to increase the
signal statistics. For the sake of convenience, we choose the
normalisation of the obtained profiles for each of the three energy
bands of $4.5-6$ GeV, $6-10$ GeV, and $>10$ GeV by equating the sum
of the computed values for the five annular bins to 100. The derived
PSF profiles are shown in Table \ref{profile}.

To validate the usage of the approximate PSF for the search for
pair halos presented in Sect. 3.2, we verify that the PSF obtained
using the method developed in the current Section is similar to that
obtained by stacking spatial distributions of \grays{} recorded
during AGN flares as done in Sect. 3.1 (see Table \ref{Tab3}).
Both the reconstructed PSFs are shown in Table \ref{profile}.
We checked and found that these reconstructed PSFs agree within
the statistical error bars.  As further validation, we re-wrote the
stacked distributions of photons (after subtraction of a background)
for quiescent states for each selected redshift interval (see Table
A4) in the form of Table 4 and found that the PSF profiles agree
with the obtained stacked distributions of photons for quiescent
states within error bars. We also re-calculated the upper limits for
the fraction of gamma rays attributable to a pair halo component
using the PSFs obtained in the current Section and found that the
re-calculated upper limits are close to those reported in Sect. 3.2
\footnote{The re-calculated 95\% upper limits for the fraction
of gamma rays from the stacked AGN sample attributable to a pair
halo component + a background in the first annular bin for the
redshift interval of $z<0.2$, equal 16\% for the energy band of
4.5-6 GeV, 9\% for the energy band of 6-10 GeV, and 7\% for the
energy band of $>$10 GeV. For the redshift interval of $0.2<z<0.6$,
we found that these upper limits equal 14\% for the energy band of
4.5-6 GeV, 10\% for the energy band of 6-10 GeV, and 10\% for the
energy band of $>$10 GeV. For the redshift interval of $0.6<z<1.3$,
we found that these upper limits equal 17\% for the energy band of
4.5-6 GeV, 19\% for the energy band of 6-10 GeV, and 12\% for the
energy band of $>$10 GeV. Finally, for the redshift interval of
$1.3<z$, we found that these upper limits equal 23\% for the energy
band of 4.5-6 GeV, 21\% for the energy band of 6-10 GeV, and 16\%
for the energy band of $>$10 GeV. Therefore these facts validate the
usage of the approximate PSFs for the searches for pair halos
presented in Sect. 3.2 and the results of these searches including
the upper limits on the fraction of photons attributable to a pair
halo component.}.

\begin{table}
\centering \caption{The PSF profiles derived from the analysis of
158 flaring sources by using the method developed in Sect. 4.2 and
from only flares (see Sect. 3.1), and from the Crab pulsar
data}
\begin{tabular}{ | c | c | c | c |}
\hline
& \multicolumn{2}{r}{Energy bands} \\
\cline{2-4}
Annular bin & $4.5-6$ GeV & $6-10$ GeV & $>10$ GeV\\
\hline
Method from Sect. 4.2 & -  & - & - \\
$0^{\circ}<$r$<0^{\circ}.21$ & $61.8\pm3.8$ & $69.8\pm3.9$ & $79.4\pm4.8$ \\
$0^{\circ}.21<$r$<0^{\circ}.3$ & $19.6\pm2.1$ & $14.6\pm1.7$ & $9.9\pm1.7$ \\
$0^{\circ}.3<$r$<0^{\circ}.42$ & $11.8\pm1.6$ & $9.1\pm1.1$ & $6.5\pm1.4$ \\
$0^{\circ}.42<$r$<0^{\circ}.52$ & $3.7\pm1.0$ & $5.0\pm0.6$ & $3.4\pm0.9$ \\
$0^{\circ}.52<$r$<0^{\circ}.6$ & $3.0\pm0.8$ & $1.6\pm0.6$ & $0.8\pm0.5$ \\
\hline
Method from Sect. 3.1 & -  & - & - \\
$0^{\circ}<$r$<0^{\circ}.21$ & $60.9\pm2.6$ & $68.5\pm2.6$ & $78.8\pm3.2$ \\
$0^{\circ}.21<$r$<0^{\circ}.3$ & $19.0\pm1.5$ & $15.9\pm1.3$ & $8.9\pm1.1$ \\
$0^{\circ}.3<$r$<0^{\circ}.42$ & $13.0\pm1.2$ & $9.3\pm1.0$ & $6.6\pm1.0$ \\
$0^{\circ}.42<$r$<0^{\circ}.52$ & $4.4\pm0.7$ & $4.3\pm0.7$ & $3.8\pm0.8$ \\
$0^{\circ}.52<$r$<0^{\circ}.6$ & $2.6\pm0.6$ & $2.0\pm0.5$ & $1.8\pm0.6$ \\
\hline
From the Crab pulsar data & -  & - & - \\
$0^{\circ}<$r$<0^{\circ}.21$ & $57.2\pm2.1$ & $67.0\pm2.2$ & $81.2\pm2.4$ \\
$0^{\circ}.21<$r$<0^{\circ}.3$ & $17.9\pm1.1$ & $15.4\pm1.1$ & $9.2\pm0.8$ \\
$0^{\circ}.3<$r$<0^{\circ}.42$ & $14.8\pm1.1$ & $11.4\pm0.9$ & $5.8\pm0.7$ \\
$0^{\circ}.42<$r$<0^{\circ}.52$ & $6.7\pm0.7$ & $3.4\pm0.5$ & $2.3\pm0.4$ \\
$0^{\circ}.52<$r$<0^{\circ}.6$ & $3.4\pm0.5$ & $2.7\pm0.5$ & $1.4\pm0.4$ \\
\hline
\end{tabular}
\label{profile}
\end{table}

\section{Comparison of methods to reconstruct the Fermi-LAT PSF}

\begin{figure}
\centering
\includegraphics[angle=0,width=8.0cm]{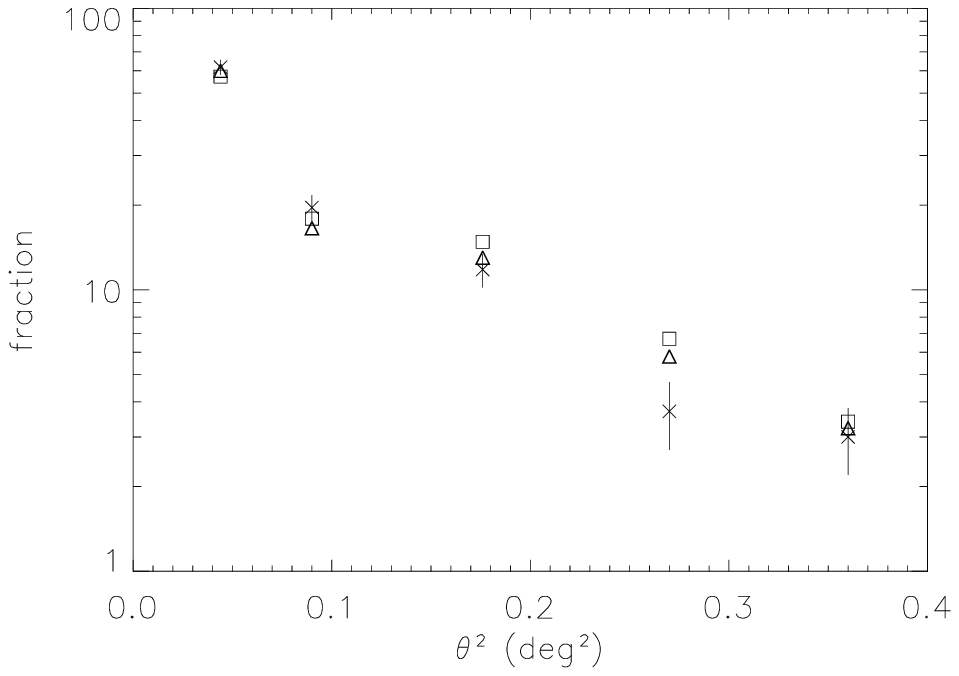}\\~
\\
\includegraphics[angle=0, width=8.0cm]{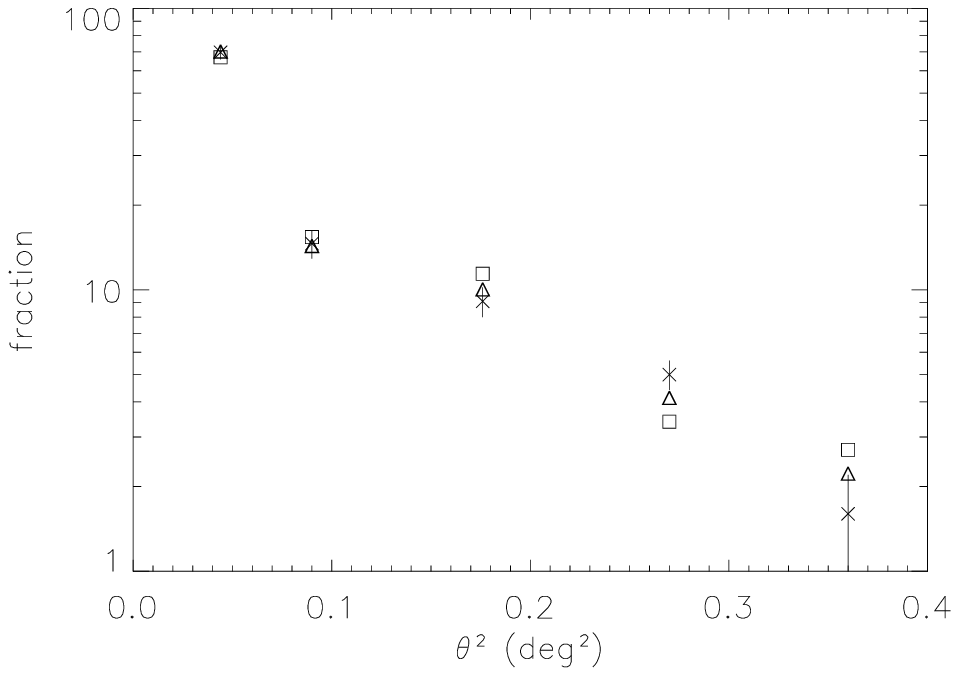}\\~
\\
\includegraphics[angle=0, width=8.0cm]{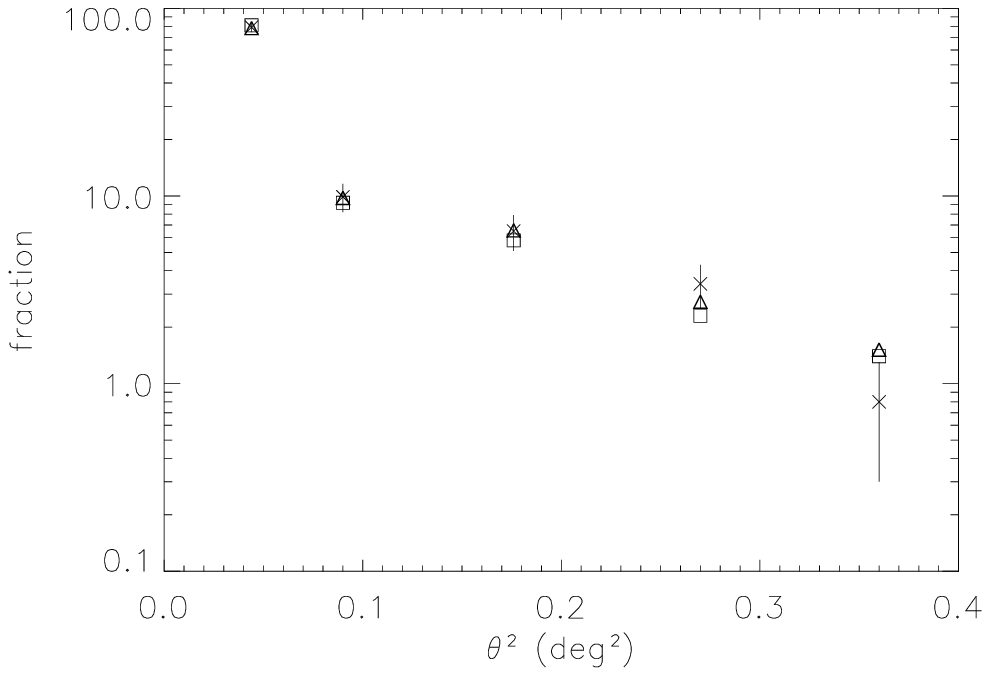}
\caption{The best-fit values of fractions (the sum of the fractions
over the five annular bins equals 100) obtained by modelling
(triangles), the observed values with error bars computed by using
AGN flares (crosses), and the central values of fractions obtained
from the observations of Crab (squares). The upper panel is for the
energy band of $4.5-6$ GeV, the middle panel is for the energy band
of $6-10$ GeV, and the bottom panel is for $>10$ GeV.} \label{F1}
\end{figure}

In this Section, first we compare the PSF computed by using the
method developed in Sect. 4 with the PSF provided by the \fermilat{}
collaboration for an analysis of Pass 7 Reprocessed Fermi dataset.
Furthermore, we compare the PSF computed above with that obtained by
using the Crab pulsar and nebula.

\subsection{Released P7REP\_SOURCE IRF PSF}

In the P7REP\_SOURCE\_V15 instrument response functions (IRFs) the
PSF is derived from flight data and are provided by the \fermilat{}
team. The updated instrument calibrations used to produce the P7REP
data release improve the LAT PSF above 3 GeV, resulting in a better
overall agreement between the Monte Carlo (MC) PSF model and the
angular distributions of gamma-ray point sources. However the MC PSF
was still found to slightly underestimate the PSF width above 3
GeV\footnote{\burl{http://fermi.gsfc.nasa.gov/ssc/data/analysis/LAT_caveats.html}}.
The LAT team has derived a new on-orbit PSF for P7REP data (included
in the P7REP\_V15 IRFs) by rescaling the MC PSF model to match the
angular distribution of Vela below 10 GeV and the stacked
distribution of bright, high-latitude blazars above 10 GeV. In order
to provide an independent validation of the P7REP\_SOURCE PSF for an
analysis of point sources with the Fermi Science
Tools\footnote{\burl{http://fermi.gsfc.nasa.gov/ssc/data/analysis/}},
we perform a comparison of the PSF derived in Sect. 4.2 with the
P7REP\_SOURCE PSF released by the LAT team.

Using the routine, \textit{gtmodel}, we model a point source using
the P7REP\_SOURCE\_V15 IRFs for three energy bands, $4.5-6$, $6-10$,
and $>10$ GeV. To compare the PSF derived in Sect. 4.2 with that
modelled for a point source, we compute a normalisation factor for a
modelled point source by minimising the $\chi^2$ value. Note that we
use the 5 annular bins described in Sect. 3. In Fig.\ref{F1}, we
show the best-fit values of fractions (the sum of the fractions over
the five annular bins equals 100) obtained by modelling (triangles)
and the observed values with error bars computed by using the method
developed in Sect. 4. The observed and modelled distributions look
very similar in Fig.\ref{F1}. To quantitatively assess the quality
of the best fits, we computed the $\chi^2$ values and found the
$\chi^2$ values equal 7.4, 3.9, and 2.6 at 4 d.o.f. for the three
energy bands, $4.5-6$, $6-10$, and $>10$ GeV, respectively.
Therefore, no significant deviations of the modelled spatial
distribution from the spatial distribution computed by using the
procedure developed in Sect. 4 are found. We conclude that the
method of the PSF reconstruction developed in Sect. 4 independently
validates the usage of the P7REP\_SOURCE PSF.

\subsection{PSF from the Crab pulsar data}

The \gray{} signal from the bright galactic gamma-ray source at the location of the Crab pulsar
in the Fermi energy band consists of two contributions: emission from the pulsar and from
the associated pulsar wind nebula (PWN). \citet[][]{Neronov2011} noticed that the size of
the associated PWN is below the angular resolution of the LAT telescope onboard Fermi and proposed the
Crab pulsar for the \fermilat{} PSF calibration.

We applied the same data reduction procedures to the data for the
region of the Crab pulsar and accumulated events for the same time
interval described in Sect. 2. The distribution of photons in
annular bins for three energy bands are presented in Table
\ref{tabCrab}. Note that the Crab pulsar provides us with a higher
number of photons than that provided by AGN flares. \gray{} flares
computed above for our sample of AGNs and accumulated during 5.2
years provide us with the number of photons which corresponds to
$\simeq3.5$ years of the Crab pulsar observation with \fermilat{}.
In Table \ref{profile}, we show the PSF profiles derived from the
Crab pulsar data (the sum of the fractions over the five annular
bins for each energy band equals 100). We checked and found that the
PSF profiles derived from the Crab pulsar data agree well with those
that are derived by using the method developed in Sect. 4. The
future observations with \fermilat{} are important for a further
comparison of the PSFs which are reconstructed with these two
different methods. Note that 10 years of \fermilat{} observations
and better event reconstruction algorithms (resulting in a higher
effective area) are expected to make error bars tighter by a factor
of $\approx$2.

\section{Conclusion}

We developed a method to search for pair halos through a temporal
analysis of AGN emission. Our method is based on an analysis of the
spatial distributions of photons coming from AGN flares and from AGN
quiescent states, and a further comparison of these two spatial
distributions.

To apply the method to the \fermilat{} data, we selected 394 AGNs
for our study. We found flaring activity in 158 AGNs of our sample.
Performing the stacking of the selected AGNs, we found that the
ratios of the number of photons coming from AGNs with known
redshifts (sorted in the four redshift intervals) and detected
during quiescent states to the number of photons coming from AGNs
and detected during flares in annular bins are within the error bars
for each of the three energy bands ($4.5-6$ GeV, $6-10$ GeV, and
$>10$ GeV) and, therefore, no evidence for a pair halo component is
found in this search. We also set the 95\% upper limit for the
fraction of \grays{} from the stacked AGN quiescent states for each
redshift interval attributable to a pair halo component in the
annular bin of $r<0.21^{\circ}$.

We presented our method for the reconstruction of the \fermilat{}
PSF by using the observations of AGN flares and quiescent states
(see Sect. 4). We compare the computed PSF with that obtained by
using the \fermilat{} dataset for the Crab pulsar. We found that
both the PSFs are in good agreement. Furthermore, we applied the
method of the PSF reconstruction developed in this paper to
independently validate the usage of the P7REP\_SOURCE IRF PSF
released by the \fermilat{} team.

\section*{Acknowledgements}

Computations were performed using a high performance computing
cluster in the Korea Astronomy and Space Science Institute. We are
grateful to the referees for the constructive suggestions that
helped us to improve this manuscript.

\appendix

\section{}

The stacked distributions of photons in annular bins for flares and
for quiescent states for each selected  energy band are shown in
Table \ref{Tab2}. The calculated numbers of photons in annular bins
for flares and for quiescent states for each selected energy band
after subtraction of a background are shown in Table \ref{Tab3}. The
ratios of the number of photons coming from 394 AGNs and detected
during quiescent states to the number of photons coming from AGNs
and detected during flares are shown in Table \ref{Tab4}. The
stacked distributions of photons in annular bins for quiescent
states for each selected energy band and for each selected redshift
interval are shown in Table \ref{TabStackRedshift}. The observed
numbers of photons with energies between 4.5 GeV and 6 GeV in the
first two annular bins for flares and for quiescent states for the
five brightest sources from our final sample are shown in Table
\ref{Tab5}. The distributions of photons in annular bins for three
energy bands for the Crab pulsar are shown in Table \ref{tabCrab}.

\begin{table*}
\centering \caption{The stacked distributions of photons in annular
bins for flares and for quiescent states for each selected  energy
band.}
\begin{tabular}{ | c | c | c | c | c |}
\hline
& \multicolumn{3}{r}{Energy bands} \\
\cline{3-5}
Annular bin & State & $4.5-6$ GeV & $6-10$ GeV & $>10$ GeV\\
\hline
$0^{\circ}<$r$<0^{\circ}.21$ & flare & 574 & 686 & 620 \\
$0^{\circ}.21<$r$<0^{\circ}.3$ & flare & 182 & 162 & 73 \\
$0^{\circ}.3<$r$<0^{\circ}.42$ & flare & 130 & 101 & 58 \\
$0^{\circ}.42<$r$<0^{\circ}.52$ & flare & 49 & 51 & 36 \\
$0^{\circ}.52<$r$<0^{\circ}.6$ & flare & 32 & 28 & 20 \\
$1^{\circ}.0<$r$<1^{\circ}.3$ & flare & 58 & 61 & 49\\
\hline
$0<$r$<0^{\circ}.21$ & quiescent & 3040 & 3881 & 5019 \\
$0^{\circ}.21<$r$<0^{\circ}.3$ & quiescent & 1012 & 992 & 802 \\
$0^{\circ}.3<$r$<0^{\circ}.42$ & quiescent & 920 & 1001 & 889 \\
$0^{\circ}.42<$r$<0^{\circ}.52$ & quiescent & 702 & 675 & 644\\
$0^{\circ}.52<$r$<0^{\circ}.6$ & quiescent & 521 & 573 & 527\\
$1^{\circ}.0<$r$<1^{\circ}.3$ & quiescent & 3220 & 3538 & 3287\\
\hline
\end{tabular}
\label{Tab2}
\end{table*}

\begin{table*}
\centering \caption{The calculated numbers of photons in annular
bins for flares and for quiescent states for each selected  energy
band after subtraction of a background.}
\begin{tabular}{ | c | c | c | c | c |}
\hline
& \multicolumn{3}{r}{Energy bands} \\
\cline{3-5}
Annular bin & State & $4.5-6$ GeV & $6-10$ GeV & $>10$ GeV\\
\hline
$0^{\circ}<$r$<0^{\circ}.21$ & flare & 570$\pm$24 & 682$\pm$26 & 617$\pm$25 \\
$0^{\circ}.21<$r$<0^{\circ}.3$ & flare & 178$\pm$14 & 158$\pm$13 & 70$\pm$9 \\
$0^{\circ}.3<$r$<0^{\circ}.42$ & flare & 122$\pm$11 & 93$\pm$10 & 52$\pm$8 \\
$0^{\circ}.42<$r$<0^{\circ}.52$ & flare & 41$\pm$7 & 43$\pm$7 & 30$\pm$6 \\
$0^{\circ}.52<$r$<0^{\circ}.6$ & flare & 24$\pm$6 & 20$\pm$5 & 14$\pm$5 \\
\hline
$0<$r$<0^{\circ}.21$ & quiescent & 2830$\pm$55 & 3650$\pm$62 & 4805$\pm$71 \\
$0^{\circ}.21<$r$<0^{\circ}.3$ & quiescent & 802$\pm$32 & 761$\pm$32 & 588$\pm$29 \\
$0^{\circ}.3<$r$<0^{\circ}.42$ & quiescent & 500$\pm$30 & 540$\pm$32 & 460$\pm$31 \\
$0^{\circ}.42<$r$<0^{\circ}.52$ & quiescent & 282$\pm$26 & 214$\pm$27 & 216$\pm$26\\
$0^{\circ}.52<$r$<0^{\circ}.6$ & quiescent & 101$\pm$23 & 112$\pm$25 & 98$\pm$24\\
\hline
\end{tabular}
\label{Tab3}
\end{table*}

\begin{table*}
\centering \caption{The ratios of the number of photons coming from
394 AGNs and detected during quiescent states to the number of
photons coming from AGNs and detected during flares}
\begin{tabular}{ | c | c | c | c |}
\hline
& \multicolumn{2}{r}{Energy bands} \\
\cline{2-4}
Annular bin & $4.5-6$ GeV & $6-10$ GeV & $>10$ GeV\\
\hline
$0^{\circ}<$r$<0^{\circ}.21$ & $5.0\pm0.2$ & $5.4\pm0.2$ & $7.8\pm0.3$ \\
$0^{\circ}.21<$r$<0^{\circ}.3$ & $4.5\pm0.4$ & $4.8\pm0.4$ & $8.4\pm1.1$ \\
$0^{\circ}.3<$r$<0^{\circ}.42$ & $4.1\pm0.5$ & $5.8\pm0.7$ & $8.9\pm1.5$ \\
$0^{\circ}.42<$r$<0^{\circ}.52$ & $6.9\pm1.3$ & $5.0\pm1.0$ & $7.3\pm1.7$ \\
$0^{\circ}.52<$r$<0^{\circ}.6$ & $4.2\pm1.4$ & $5.6\pm2.0$ & $7.2\pm3.0$ \\
\hline
$0^{\circ}.21<$r$<0^{\circ}.6$ & $4.5\pm0.3$ & $5.0\pm0.3$ & $8.2\pm0.8$ \\
\hline
\end{tabular}
\label{Tab4}
\end{table*}

\begin{table*}
\centering \caption{The stacked distributions of photons in annular
bins for quiescent states for each selected energy band and for each
selected redshift interval.}
\begin{tabular}{ | c | c | c | c | c |}
\hline
& \multicolumn{3}{r}{Energy bands} \\
\cline{3-5}
Redshift & Annular bin & $4.5-6$ GeV & $6-10$ GeV & $>10$ GeV\\
\hline
$z<0.2$ & $0^{\circ}<$r$<0^{\circ}.21$ & 504 & 752 & 1346 \\
$z<0.2$ & $0^{\circ}.21<$r$<0^{\circ}.3$ & 165 & 163 & 181 \\
$z<0.2$ & $0^{\circ}.3<$r$<0^{\circ}.42$ & 118 & 132 & 165 \\
$z<0.2$ & $0^{\circ}.42<$r$<0^{\circ}.52$ & 79 & 89 & 104 \\
$z<0.2$ & $0^{\circ}.52<$r$<0^{\circ}.6$ & 47 & 50 & 57 \\
$z<0.2$ & $1^{\circ}.0<$r$<1^{\circ}.3$ & 294 & 339 & 327 \\
\hline
$0.2<z<0.6$ & $0^{\circ}<$r$<0^{\circ}.21$ & 546 & 666 & 836 \\
$0.2<z<0.6$ & $0^{\circ}.21<$r$<0^{\circ}.3$ & 173 & 147 & 133 \\
$0.2<z<0.6$ & $0^{\circ}.3<$r$<0^{\circ}.42$ & 150 & 168 & 131 \\
$0.2<z<0.6$ & $0^{\circ}.42<$r$<0^{\circ}.52$ & 120 & 113 & 110 \\
$0.2<z<0.6$ & $0^{\circ}.52<$r$<0^{\circ}.6$ & 78 & 102 & 94 \\
$0.2<z<0.6$ & $1^{\circ}.0<$r$<1^{\circ}.3$ & 461 & 472 & 497 \\
\hline
$0.6<z<1.3$ & $0<$r$<0^{\circ}.21$ & 639 & 784 & 765 \\
$0.6<z<1.3$ & $0^{\circ}.21<$r$<0^{\circ}.3$ & 216 & 227 & 135 \\
$0.6<z<1.3$ & $0^{\circ}.3<$r$<0^{\circ}.42$ & 202 & 197 & 154 \\
$0.6<z<1.3$ & $0^{\circ}.42<$r$<0^{\circ}.52$ & 149 & 138 & 121\\
$0.6<z<1.3$ & $0^{\circ}.52<$r$<0^{\circ}.6$ & 107 & 132 & 104\\
$0.6<z<1.3$ & $1^{\circ}.0<$r$<1^{\circ}.3$ & 703 & 741 & 681\\
\hline
$1.3<z$ & $0<$r$<0^{\circ}.21$ & 333 & 311 & 271 \\
$1.3<z$ & $0^{\circ}.21<$r$<0^{\circ}.3$ & 121 & 89 & 51 \\
$1.3<z$ & $0^{\circ}.3<$r$<0^{\circ}.42$ & 120 & 116 & 88 \\
$1.3<z$ & $0^{\circ}.42<$r$<0^{\circ}.52$ & 98 & 95 & 68\\
$1.3<z$ & $0^{\circ}.52<$r$<0^{\circ}.6$ & 76 & 74 & 58\\
$1.3<z$ & $1^{\circ}.0<$r$<1^{\circ}.3$ & 464 & 476 & 438\\
\hline
\end{tabular}
\label{TabStackRedshift}
\end{table*}

\begin{table*}
\centering \caption{The observed numbers of photons with energies
between 4.5 GeV and 6 GeV in the first two annular bins for flares
and for quiescent states for the five brightest sources from our
final sample.}
\begin{tabular}{ | c | c | c | c | c | c |}
\hline
& \multicolumn{3}{r}{Annular bins} \\
\cline{4-6}
Source name & State & Energy band & $0^{\circ}<$r$<0^{\circ}.21$ &
$0^{\circ}.21<$r$<0^{\circ}.42$ & Number of intervals\\
\hline
4C +21.35 & flare & 4.5-6 GeV & 51 & 34 & 60 \\
3C 279 & flare & 4.5-6 GeV & 21 & 13 & 40 \\
PKS 1510-08 & flare & 4.5-6 GeV & 56 & 22 & 80 \\
PKS 0537-441 & flare & 4.5-6 GeV & 34 & 16 & 51 \\
Mkn 421 & flare & 4.5-6 GeV & 17 & 7 & 20 \\
\hline
4C +21.35 & quiescent & 4.5-6 GeV & 32 & 21 & 633 \\
3C 279 & quiescent & 4.5-6 GeV & 49 & 21 & 497 \\
PKS 1510-08 & quiescent & 4.5-6 GeV & 73 & 44 & 1052 \\
PKS 0537-441 & quiescent & 4.5-6 GeV & 106 & 48 & 858\\
Mkn 421 & quiescent & 4.5-6 GeV & 228 & 110 & 1206\\
\hline
4C +21.35 & flare & 6-10 GeV & 65 & 22 & 60 \\
3C 279 & flare & 6-10 GeV & 17 & 9 & 40 \\
PKS 1510-08 & flare & 6-10 GeV & 49 & 20 & 80 \\
PKS 0537-441 & flare & 6-10 GeV & 62 & 17 & 51 \\
Mkn 421 & flare & 6-10 GeV & 36 & 9 & 20 \\
\hline
4C +21.35 & quiescent & 6-10 GeV & 42 & 10 & 633 \\
3C 279 & quiescent & 6-10 GeV & 36 & 21 & 497 \\
PKS 1510-08 & quiescent & 6-10 GeV & 83 & 32 & 1052 \\
PKS 0537-441 & quiescent & 6-10 GeV & 140 & 46 & 858 \\
Mkn 421 & quiescent & 6-10 GeV & 364 & 133 & 1206 \\
\hline
\end{tabular}
\label{Tab5}
\end{table*}

\begin{table*}
\centering \caption{The distributions of photons in annular bins for
three energy bands for the Crab pulsar.}
\begin{tabular}{ | c | c | c | c | c |}
\hline
& \multicolumn{3}{r}{Energy bands} \\
\cline{3-5}
Quantity & Annular bin & $4.5-6$ GeV & $6-10$ GeV & $>10$ GeV\\
\hline
Photons & $0^{\circ}<$r$<0^{\circ}.21$ & 782 & 948 & 1118 \\
Photons & $0^{\circ}.21<$r$<0^{\circ}.3$ & 247 & 220 & 129 \\
Photons & $0^{\circ}.3<$r$<0^{\circ}.42$ & 208 & 164 & 85 \\
Photons & $0^{\circ}.42<$r$<0^{\circ}.52$ & 98 & 51 & 37 \\
Photons & $0^{\circ}.52<$r$<0^{\circ}.6$ &  54 & 41 & 24 \\
Photons & $1^{\circ}.0<$r$<1^{\circ}.3$ & 56 & 35 & 37\\
\hline
\end{tabular}
\label{tabCrab}
\end{table*}

\bibliography{refs}

\end{document}